\documentclass[a4paper,10pt]{article}

\usepackage[max2]{authblk}

\usepackage{hyperref}
\usepackage{amsmath}
\usepackage{amssymb}

\usepackage{blkarray}
\usepackage{fullpage}

\usepackage{multirow}

\usepackage{cite}

\usepackage{graphicx}
\usepackage{subfigure}

\usepackage{xcolor}

\xdefinecolor{darkblue}{rgb}{0.,0.,0.7}
\xdefinecolor{darkgreen}{rgb}{0,0.43,0}
\xdefinecolor{darkorange}{rgb}{0.8,0.44,0}

\title{\begin{flushright}\large
\vspace{-1cm} DESY 14-042
\end{flushright}\vspace{1cm}\textbf{Towards discrimination of MSSM and NMSSM scenarios at colliders}}
\author[1]{Stefano Porto\footnote{Speaker, \texttt{stefano.porto@desy.de}}}
\author[1,2]{Gudrid A. Moortgat-Pick\footnote{\texttt{gudrid.moortgat-pick@desy.de}}}
\author[3]{Krzysztof Rolbiecki\footnote{\texttt{krzysztof.rolbiecki@desy.de}}}
\affil[1]{ \small II. Institut f\"{u}r Theoretische Physik, University of Hamburg, Germany\normalsize}
\affil[2]{\small DESY, Hamburg, Germany\normalsize}
\affil[3]{ \small  Instituto de Fisica Teorica IFT-UAM/CSIC, Universidad Autonoma de Madrid, Spain\normalsize}

\begin{document}

\date{Talk presented at the International Workshop on Future Linear Colliders (LCWS13), Tokyo, Japan, 11-15 November 2013.}

\maketitle

\begin{abstract}
One of the challenging tasks at future experiments is the clear
identification of the underlying new physics model.  
In this study we concentrate on the distinction between
different supersymmetric models, the MSSM and the NMSSM,
 exploring the gaugino/higgsino sector
as an alternative to the Higgs sector.  Under the assumption that only
the light chargino and neutralino masses and polarized cross sections
$e^+e^-\to \tilde{\chi}^0_i\tilde{\chi}^0_j$,
$\tilde{\chi}^+_i\tilde{\chi}^-_j$ have been measured, we perform a fit of
the fundamental MSSM parameters $M_1$, $M_2$, $\mu$ and
$\tan\beta$ and study whether a model distinction is possible.  We focus
here on the challenging cases of scenarios with a relatively heavy
singlino and address two classes of neutralino mixing,
$\tilde{\chi}^0_1\sim$higgsino-like versus
$\tilde{\chi}^0_1\sim$gaugino-like.
\end{abstract}

\section{Introduction}
Supersymmetry is an appealing candidate for  physics Beyond the
Standard Model (BSM): the introduction of a (broken) fermion-boson
symmetry answers elegantly, for instance,
the electroweak hierarchy puzzle,
offers a dark matter candidate and is consistent with grand unification.  The
recent discovery of a Higgs-like particle with $m_h \sim 125.5$ GeV at the
LHC \cite{Aad:2012tfa,Chatrchyan:2012ufa}, together with the negative result of SUSY searches, 
however, is posing a 
challenge for theorists to conciliate supersymmetry.

Well motivated candidates for BSM are the Minimal Supersymmetric
Standard Model (MSSM) and its minimal extension, the next-to-minimal
Standard Model (NMSSM). The latter adds a gauge singlet chiral supermultiplet
$\hat{S}$ and allows therefore 
a relaxation of the electroweak fine tuning and the
naturalness conditions.

It is interesting to understand how, in case of a discovery of
supersymmetry 
at the LHC and at a future linear collider, it would be possible to
discriminate between these two models. They have, indeed, a very similar
particle content but the NMSSM offers
an enriched Higgs and higgsino spectrum:  
a CP-even Higgs, a CP-odd Higgs and a fifth neutralino in addition to the
MSSM. In the
literature mainly NMSSM scenarios are studied that
have a singlino-like stable lightest supersymmetric particle (LSP).  
Due to the expected experimental accuracy in the
Higgs measurements both at the LHC and the ILC
\cite{Dawson:2013bba,Asner:2013psa}, and the additional two scalar 
states, a standard procedure to pinpoint the observed supersymmetry
model is to study the Higgs sector.  However, it could also well be 
that the additional Higgs bosons, are
very heavy and not clearly detectable at the LHC/ILC. 
It is therefore important to find
alternative and complementary tools to distinguish these models, 
for example addressing also the gaugino/higgsino sector.

We follow such an alternative ansatz and assume that only the
lightest states in this sector $\tilde{\chi}^0_1$, $\tilde{\chi}^0_2$
and $\tilde{\chi}^{\pm}_1$ are accessible, however contrary to
\cite{MoortgatPick:2005vs}, we focus on the heavy singlino case.
Given the masses and production cross sections measured with a precision
expected to be achievable at the linear collider, this
method reconstructs the MSSM chargino and neutralino sector parameters
$M_1$, $M_2$, $\mu=\mu_{\rm eff}$, $\tan\beta$ in
the philosophy of \cite{Desch:2003vw}. We perform
a $\chi^2$-fits and study whether a result 
non-compatible with the MSSM can be the
smoking gun for the NMSSM. 

In this proceeding, we shortly introduce the subject of an upcoming
paper \cite{MoortgatPick:2014}. In particular, we address NMSSM
scenarios with relatively heavy singlino ($\tilde{S}$), such that the
detected spectra could be interpreted as an MSSM signal.  We scan the
$(\lambda,\,\kappa)$-plane applying the most recent phenomenological and
experimental constraints from colliders and dark matter experiments in
two cases. First we study a scenario with higgsino-like LSP and
relatively heavy gauginos; then we look at a scenario with wino-like LSP
($M_1>M_2$), expected in the context of minimal AMSB models
\cite{Randall:1998uk,Giudice:1998xp}.

In section \ref{Sec_Strategy} we briefly describe the strategy to
discriminate the models; in section \ref{Sec_Scenarios} we show the
results of our analysis for different scenarios before we shortly summarize
in section \ref{Sec_Conclusions}.

\section{Strategy and MSSM parameter reconstruction}\label{Sec_Strategy}
 The $\mathbb{Z}_3$-invariant NMSSM, with the additional term in the superpotential
\begin{equation}
 W_{\mbox{\scriptsize NMSSM\normalsize}}=\lambda\,\hat{S}\hat{H}_u\cdot\hat{H}_d+\frac{\kappa}{3}\,\hat{S}^3\,,
\end{equation}
features, with respect to the MSSM, a further gauge singlet superfield
$\hat{S}$, consisting of a scalar Higgs singlet $S$ and a neutralino
$\tilde{S}$ that mixes due to electroweak symmetry breaking with the gaugino/higgsinos states.
Therefore, looking for weakly coupling scalars or neutralinos is
na\"{\i}vely the first way to discriminate between NMSSM and
MSSM, in particular promising
in the light of the expected high accuracy in the Higgs
sector measurements \cite{Asner:2013psa}.

However, in case that the singlet states
are relatively heavy in comparison with the SM-like Higgs, a
discrimination could be more challenging since the
observed Higgs sector can be interpreted within both the MSSM and the
NMSSM. Signal strengths at the LHC would be very similar in both models with the heavier states 
decoupled from the spectrum and beyond the kinematic reach at the future linear
collider.

The MSSM chargino and neutralino sectors are fully described by the
 parameters $M_1$, $M_2$, $\mu$, $\tan \beta$. It has been shown in
 \cite{Choi:2000ta,Choi:2001ww,Choi:2002mc}, that full tree-level determination of
 these parameters is possible at a linear collider, provided that 
 $\tilde{\chi}^0_1$, $\tilde{\chi}^0_2$ and $\tilde{\chi}^{\pm}_1$
 can be produced at the LC and their masses as well as the polarized
 cross sections
 $\sigma(e^+e^-\rightarrow\tilde{\chi}^0_1\tilde{\chi}^0_2)$,
 $\sigma(e^+e^-\rightarrow\tilde{\chi}^+_1\tilde{\chi}^-_1)$ are
 precisely measured.  A $\chi^2$-minimisation selects
 parameters fitting the experimental results and provides a precise
 and rather model-independent determination of $M_1$, $M_2$, $\mu$, 
$\tan \beta$. Such analysis can be strengthened if the
 mass of the heavier neutralino states can be inferred from combined
 analyses of LHC and LC data~\cite{Desch:2003vw}.
 
The possibility of reconstructing the MSSM  chargino-neutralino sector
parameters can then be enveloped in a strategy to discriminate between
the MSSM and the NMSSM \cite{MoortgatPick:2005vs}, complementary to
looking only at the Higgs sectors. 
 If the result of the $\chi^2$-fit, based only
on the measured $\tilde{\chi}^0_1$, $\tilde{\chi}^0_2$ and
 $\tilde{\chi}^{\pm}_1$ sector,
excludes the MSSM at 95\% C.L.,  
one should look for extended models as, for instance, the NMSSM.

In this study, we address in particular 
challenging NMSSM scenarios with relatively
heavy singlino (and singlet) but a lower chargino-neutralino spectra
that is approximately MSSM-like and proceed as follows:
%%%
\begin{itemize}
 \item We choose an NMSSM scenario that presents low chargino-neutralino
       spectrum that is nearly indistinguishable with
       respect to the one of a corresponding MSSM scenario.

 \item The $\lambda,\,\kappa$ parameters encode the pure NMSSM-behaviour
in the neutralino sector and change the singlino admixtures in the
       different mass eigenstates.
We scan a grid of ten thousand points in the
$(\lambda,\,\kappa)$-plane for values $\lambda\in[0,0.7]$ and
$\kappa\in[0,0.7]$. The singlino character of the neutralinos has a strong
impact on the suitable strategy for the distinction of both models.

Each point, in order to be further considered in the analysis, has to
pass a series of phenomenological and experimental constraints
implemented in \texttt{NMSSMTools-4.2.1}, that includes
\texttt{NMHDECAY}
\cite{Ellwanger:2004xm,Ellwanger:2005dv,Belanger:2005kh} and
\texttt{NMSDECAY} \cite{Das:2011dg,Muhlleitner:2003vg}. These tools
calculate the Higgs sector parameters, SUSY particle masses at the loop level and their decays
and test their agreement with limits from
LEP and LHC and other EW precision
constraints. Dark matter constraints, including the latest LUX and
Planck results, are implemented through an interface to \texttt{MicrOMEGAS}
\cite{Belanger:2013oya}. We require the LSP relic density to be $\Omega_{\rm LSP}h^2< 0.131$, where $h$ is the Hubble constant in units of 100 km/(s$\cdot$Mpc). A second test
on the Higgs sector constraints is done with \texttt{HiggsBounds-4.0.0}
\cite{Bechtle:2013wla} and \texttt{HiggsSignals-1.0.0}
\cite{Bechtle:2013xfa}, and we accept only points compatible at 95\%
C.L. with the current data.  
  \item For each point in the $(\lambda,\,\kappa)$-plane passing the
       tests mentioned above, we assume that the lower chargino/neutralino spectrum,
       namely $\tilde{\chi}^0_1$, $\tilde{\chi}^0_2$ and
       $\tilde{\chi}^{\pm}_1$, is observed at the ILC. We include 
the tree-level
       masses and production cross-sections for the processes
       $e^+e^-\rightarrow\tilde{\chi}^+_1\tilde{\chi}^-_1$ and
       $e^+e^-\rightarrow\tilde{\chi}^0_1\tilde{\chi}^0_2$  or $\tilde{\chi}^0_1\tilde{\chi}^0_3$ with electron
       and positron beam polarisation
       $(\mathcal{P}_{e^{-}},\mathcal{P}_{e^{+}})= (\pm0.9,\mp0.55)$
	measured at
       $\sqrt{s}=350$ GeV ($t\bar{t}$-threshold) and at $\sqrt{s}=500$ GeV. 
A precision of
       uncertainty of 0.5\% on the masses and 1\% on the cross 
sections is assumed
       \cite{AguilarSaavedra:2001rg,Baer:2013cma}.

\item For each NMSSM point, the ``measured'' masses, cross-sections and respective
       uncertainties are used to perform the MSSM parameter
       determination through the $\chi^2$-fit following the recipe in
       \cite{Desch:2003vw}, using \texttt{Minuit} \cite{James1975343},
       minimizing
\begin{equation}
\chi^2=\sum_i\left|\frac{\mathcal{O}_i-\bar{\mathcal{O}}_i}{\delta\mathcal{O}_i}\right|^2\mbox{.}
\end{equation} 
The $\mathcal{O}_i$ are the input observables, $\delta O_i$ are the
       associated experimental uncertainties and $\bar{O}_i$ are the
theoretical values of the observables on basis of the fitted
       MSSM parameters.
If for a given point the fit is not consistent with the MSSM at the 95\% C.L., it provides an
       experimental hint towards the NMSSM. In this way we can
       identify those parameter regions where the model distinction is
       possible in spite of the challenging assumption that only a very
      limited amount of experimental observables are accessible.
\end{itemize}

\begin{figure}[htb]\centering
\includegraphics[width=1\textwidth]{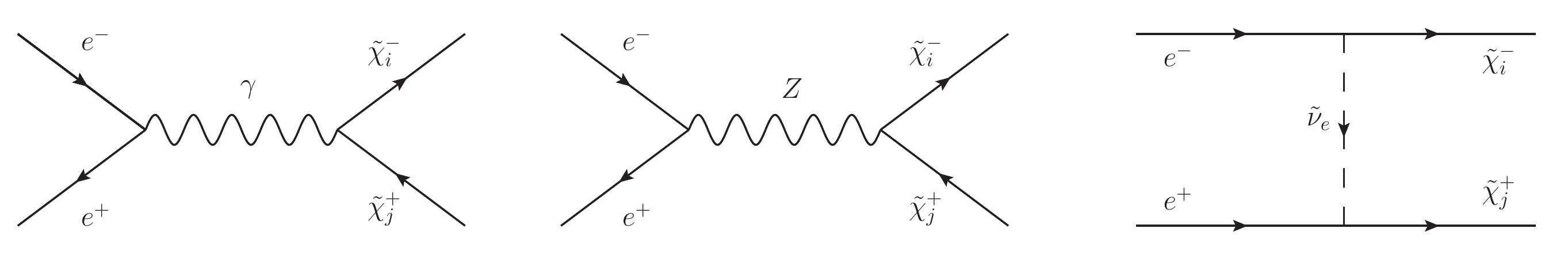}
\caption{Feyman diagrams for $\tilde{\chi}^+_i\tilde{\chi}^-_j$ production at $e^+e^-$ colliders.\label{Fig_charginoFeynman}}
\end{figure}

\begin{figure}[htb]\centering
\includegraphics[width=1\textwidth]{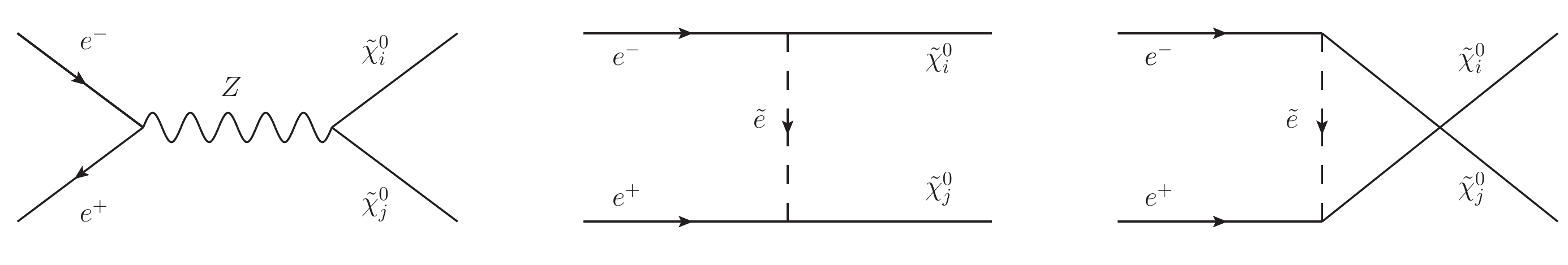}
\caption{Feynman diagrams for $\tilde{\chi}^0_i\tilde{\chi}^0_j$ production at $e^+e^-$ colliders.\label{Fig_neutralinoFeynman}}
\end{figure}

\section{Scenarios and analysis}\label{Sec_Scenarios}
 
In the NMSSM, the singlino ($\tilde{S}$) admixture of $\tilde{\chi}^0_1$  can be used to pinpoint two main classes of scenarios: 
\begin{itemize}
 \item High $\tilde{S}$ admixture in $\tilde{\chi}^{0}_{1}$ or $\tilde{\chi}^{0}_{2}$.
 
 Since we assume to detect the lightest neutralinos, a light singlino
       would be the smoking gun for a non-MSSM scenario, due to the
       different cross-sections and in consequence non-compatible
       fit.
Because of a high admixture of $\tilde{S}$ in $\tilde{\chi}^0_1$ and/or
       $\tilde{\chi}^0_2$, indeed, the higgsino and gaugino components
       in these lightest NMSSM neutralino states would be substantially
       different from those of the fit-reconstructed MSSM scenario.  In
       such cases the
       distinction between the models through the outlined method is expected
       to be promising, see \cite{MoortgatPick:2005vs}.
 
\item High $\tilde{S}$ admixture mainly in the
       heavy neutralino states
       $\tilde{\chi}^{0}_{3},\tilde{\chi}^{0}_{4},\tilde{\chi}^{0}_{5}$.

We will in particular 
focus on cases, where both the light spectra and the
       neutralino admixture are very similar between NMSSM and
       corresponding MSSM scenario. In this case it is likely that the
       fit is still compatible with the MSSM.
We question how to integrate informations 
from heavier neutralino states at the LHC or TeV-LC,
       and/or from the Higgs sector in order to enable a model distinction.

In particular we analyse an example for each of the following categories:

\normalsize\begin{enumerate}
   \item $\mu_{\rm eff}<M_1,M_2$: the LSP, $\tilde{\chi}^0_1$, is mainly
	 higgsino-like in the whole studied $(\lambda,\,\kappa)$-plane, see
	 subsection \ref{SubSecHIGLSP}.\tiny
   
 \normalsize
 \item $\mu_{\rm eff}>M_1,M_2$: $\tilde{\chi}^0_1$ is mainly 
gaugino-like in the  whole studied $(\lambda,\,\kappa)$-plane, see 
subsection \ref{SubSecGAULSP}.  \end{enumerate}  
\end{itemize}

\subsection{Light higgsino scenario, $\mu_{\rm eff}<M_1<M_2$}\label{SubSecHIGLSP}

The chargino/neutralino sector parameters for the light higgsino NMSSM scenario we consider are: 
\begin{equation}
 M_1=450\mbox{ GeV,\,\,\, }\,\,\,M_2=1600\mbox{ GeV, }\,\,\,\mu_{\rm eff}=\lambda\,s=120\mbox{ GeV, }\,\,\,\tan\beta=27\,,\label{HigLSP_parameters}
\end{equation}
while $\lambda\in[0,0.7]$ and $\kappa\in[0,0.7]$ as prescribed above and $\mu_{\rm eff}$ is kept fixed by varying $s$, the singlet vev. The singlet soft parameters are $A_{\lambda}=3000$ GeV, $A_{\kappa}=-30$ GeV. The first generation sfermion masses, needed for the production cross sections, see Figures \ref{Fig_charginoFeynman} and \ref{Fig_neutralinoFeynman}, are
\begin{equation}\label{HigLSP_sfMass}
 m_{\tilde{e}_L}=303.5\mbox{ GeV},\, m_{\tilde{e}_R}=303\mbox{ GeV},\, m_{\tilde{\nu}_e}=293.3\mbox{ GeV}\,,
\end{equation}

 while squarks masses are $>1$ TeV.

\

We now take a MSSM scenario with $M_1$, $M_2$, $\mu=\mu_{\rm eff}$, $\tan\beta$ as in \eqref{HigLSP_parameters} and the same slepton masses as in \eqref{HigLSP_sfMass}. For the lightest neutralino and chargino states we have obtained, at the tree-level, masses and  production cross sections that are very close to ones of the NMSSM scenario.
The MSSM tree-level chargino/neutralino spectrum is given by:
\begin{center}
\begin{tabular}{|c|c|c|c|c|c|}\hline
$m_{\tilde{\chi}^0_1}$&$m_{\tilde{\chi}^0_2}$&$m_{\tilde{\chi}^0_3}$&$m_{\tilde{\chi}^0_4}$&$m_{\tilde{\chi}^{\pm}_1}$&$m_{\tilde{\chi}^{\pm}_2}$\\\hline
  114.8 GeV&123.3 GeV&454.4 GeV&1604.1 GeV&119.4 GeV&1604.1 GeV\\\hline
\end{tabular}             \end{center}
\normalsize

\

\begin{figure}[htb]\centering
\subfigure[]{\label{HIGLSPN1mass}\includegraphics[width=.49\textwidth]{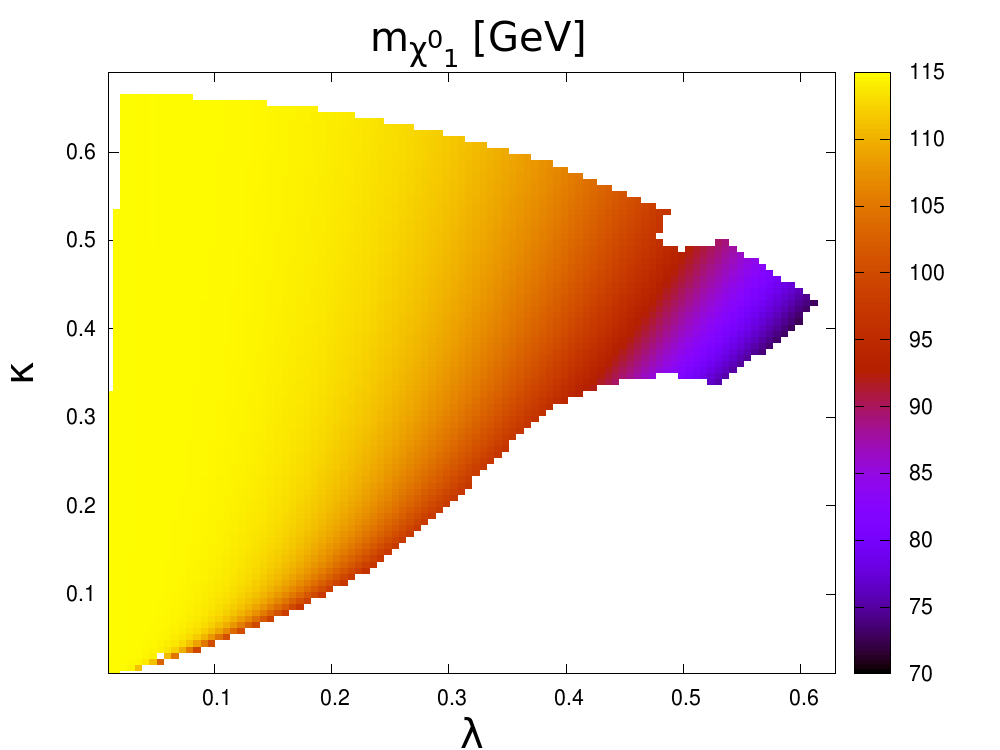}}
\subfigure[]{\label{HIGLSPN15}\includegraphics[width=.49\textwidth]{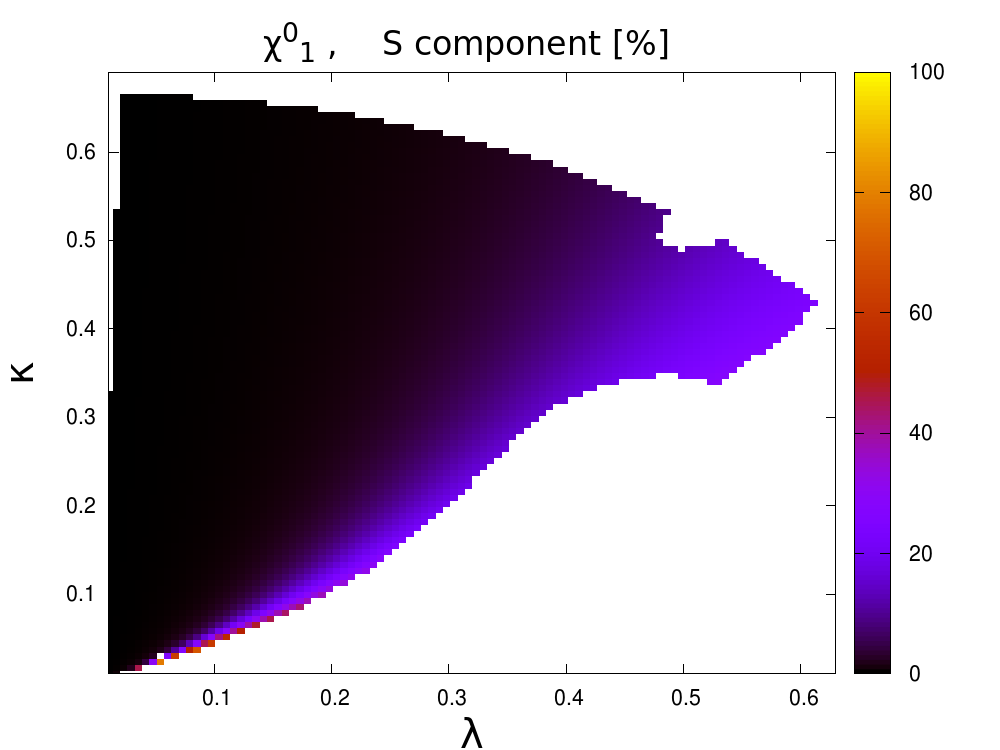}}
\caption{Light higgsino scenario: (a) the mass $m_{\tilde{\chi}^0_1}$, in GeV; (b) the $\tilde{S}$ component of $\tilde{\chi}^0_1$, in \%.\label{HIGN1plots}}
\end{figure}

In Figure \ref{HIGN1plots}, the NMSSM $\tilde{\chi}^0_1$ mass and its singlino ($\tilde{S}$) component of  are shown. One can clearly see that in the region where the singlino component is negligible, the NMSSM $m_{\tilde{\chi}^0_1}$ is very close to the MSSM value $m_{\tilde{\chi}^0_1}=114.8$. In correspondence of higher singlino admixture, instead, NMSSM $m_{\tilde{\chi}^0_1}$ sensibly lowers.

\

The polarized tree-level production cross sections for the MSSM scenario are:

\begin{center}
\begin{tabular}{|c||c|c|}\hline{\textbf{MSSM}}, $\sigma(e^+e^-\rightarrow\tilde{\chi}^0_1\tilde{\chi}^0_2)\,\,$  & $\sqrt{s}=350$~GeV & $\sqrt{s}=500$~GeV\\\hline\hline
$P=(-0.9,0.55)$&791.7 fb&391.4 fb\\\hline
$P=(0.9,-0.55)$&526.7 fb&261.7 fb\\\hline
\end{tabular} \end{center}\normalsize

\begin{center}
\begin{tabular}{|c||c|c|}\hline{\textbf{MSSM}}, $\sigma(e^+e^-\rightarrow\tilde{\chi}^+_1\tilde{\chi}^-_1)$ & $\sqrt{s}=350$~GeV & $\sqrt{s}=500$~GeV\\\hline\hline
$P=(-0.9,0.55)$ &2348.8 fb & 1218.9 fb\\\hline
$P=(0.9,-0.55)$ &445.1 fb & 246.2 fb\\\hline
\end{tabular} \end{center}

\begin{figure}[htb]\centering
\subfigure[]{\label{HIGLSPN1N2_350M9P55}\includegraphics[width=.49\textwidth]{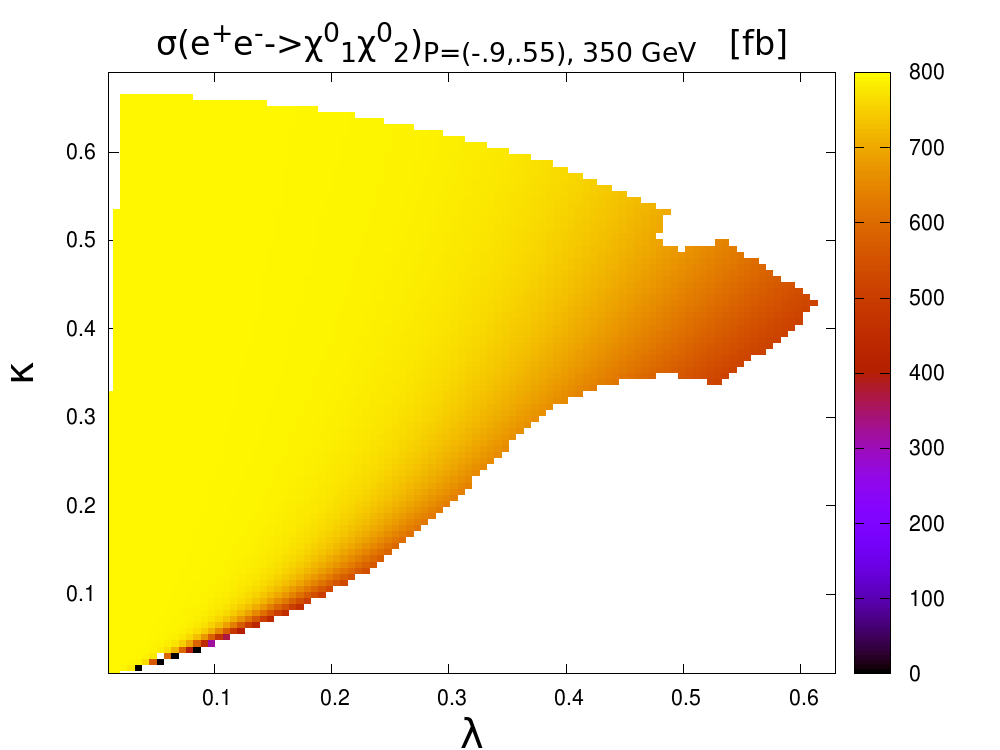}}
\subfigure[]{\label{HIGLSPN1N2_500P9M55}\includegraphics[width=.49\textwidth]{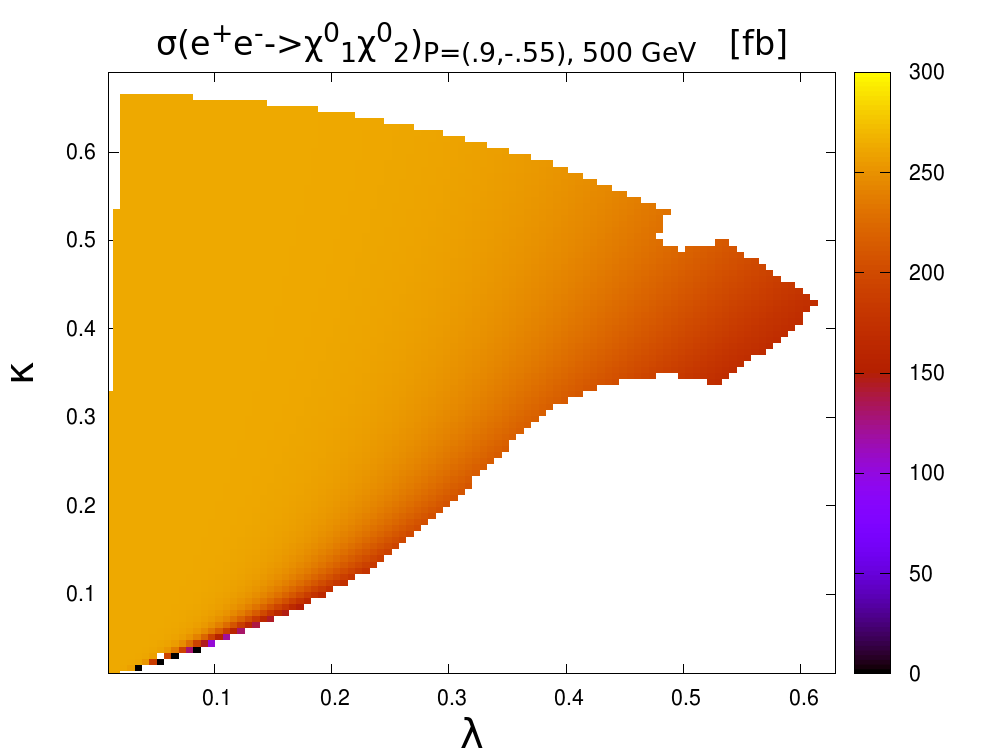}}
\caption{$\tilde{\chi}^0_1\tilde{\chi}^0_2$ production cross sections in the ight higgsino scenario: (a) $\sigma(e^+e^-\rightarrow\tilde{\chi}^0_1\tilde{\chi}^0_2)$ for $P= (-0.9,0.55)$ at $\sqrt{s} = 350$~GeV, in fb; (b) $\sigma(e^+e^-\rightarrow\tilde{\chi}^0_1\tilde{\chi}^0_2)$ for $P=(+0.9,-0.55)$ at $\sqrt{s} = 500$~GeV, in fb. \label{HIGN1N2_prod_plots}}
\end{figure}

At the tree-level, the NMSSM chargino masses and production cross-sections $\sigma(e^+e^-\rightarrow\tilde{\chi}^+_1\tilde{\chi}^-_1)$ depend only on $M_2$, $\mu_{\rm eff}$, $\tan\beta$. Therefore, in the $(\lambda,\,\kappa)$-plane chargino production cross sections are identical to those of the MSSM scenario, since this has the same $M_2$, $\mu$, $\tan\beta$ as in \eqref{HigLSP_parameters}. 
Neutralino pair production $e^+e^-\rightarrow\tilde{\chi}^0_1\tilde{\chi}^0_2$, instead, varies with $\lambda$ and $\kappa$, see Figure \ref{HIGN1N2_prod_plots}, where it can be observed a lowering of the cross section following a higher singlino component in $\tilde{\chi}^0_1$.

As described in Section \ref{Sec_Strategy} we now assume for each point in the $(\lambda,\,\kappa)$-plane experimental measurement of the corresponding lighter neutralino and chargino masses and associated production cross sections.
Next, we perform the $\chi^2$-fit to the MSSM, we assume for this scenario to detect at the ILC $m_{\tilde{\chi}^0_1}$, $m_{\tilde{\chi}^0_2}$ and $m_{\tilde{\chi}^{\pm}_1}$ with an uncertainty of $0.5\%$. Furthermore, we assume $1\%$ uncertainty on the polarized cross sections $\sigma(e^+e^-\rightarrow\tilde{\chi}^0_1\tilde{\chi}^0_2)$ and $\sigma(e^+e^-\rightarrow\tilde{\chi}^+_1\tilde{\chi}^-_1)$, both at $\sqrt{s}=350$ and 500 GeV, with polarizations $P= (-0.9,0.55)$ and $P=(+.9,-.55)$.

\begin{figure}[t!]
\begin{center}
 \includegraphics[width=3.0in]{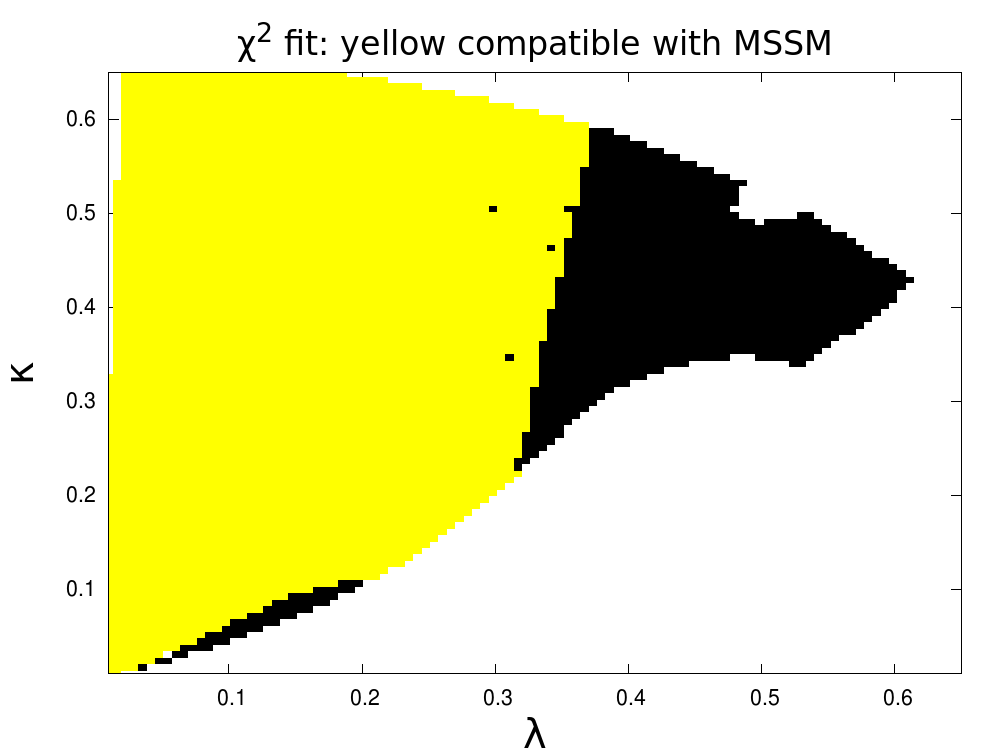}
 \caption{Light higgsino scenario: fit to the MSSM. Yellow areas are compatible with the MSSM, black areas are not compatible.\label{HigLSP_Fit}}
\end{center}
\end{figure}

In Figure \ref{HigLSP_Fit} the result of the fit is shown: yellow areas correspond to regions in the $(\lambda,\,\kappa)$-plane that are compatible with the MSSM scenario, while the black areas are not compatible. We can observe two regions that, while allowed by the implemented phenomenological and experimental constraints, can definitely be distinguished from the MSSM using collider observables.

One can ask now how to integrate this result with additional information, further reducing the region that cannot be distinguished from the MSSM scenario. % and even unambiguously pinpoint to the NMSSM as the underlying model.
This information can be recovered from the heavier neutralino states, such as $\tilde{\chi}^0_3$. For example, given a set of $M_1$, $M_2$, $\mu$, $\tan\beta$ reconstructed from the fit, one can derive $m_{\tilde{\chi}^0_3}$. Looking for such a state at the ILC or at the LHC, can either confirm the fit to the MSSM or even pinpoint the NMSSM. Further information can be given from the Higgs sector, in particular from the search of singlet states. 
Such questions will be addresses in the upcoming work \cite{MoortgatPick:2014}.

\subsection{Light gaugino scenario, $\mu_{\rm eff}>M_1>M_2$}\label{SubSecGAULSP}
We look at the light gaugino NMSSM scenario, whose neutralino/chargino sector parameters are: 
\begin{equation}
 M_1=240\mbox{ GeV,\,\,\, }\,\,\,M_2=105\mbox{ GeV, }\,\,\,\mu=\mu_{\rm eff}=505\mbox{ GeV, }\,\,\,	\tan\beta=9.2\,,\label{GauLSP_parameters}
\end{equation}
with $\lambda\in[0,0.7]$ and $\kappa\in[0,0.7]$. Moreover, $A_{\lambda}=3700$ GeV, $A_{\kappa}=-40$ GeV. The first generation sfermion masses are\begin{equation}\label{GauLSP_sfMass}
 m_{\tilde{e}_L}=303.5\mbox{ GeV},\, m_{\tilde{e}_R}=303\mbox{ GeV},\, m_{\tilde{\nu}_e}=293.3\mbox{ GeV}\,,
\end{equation} while squarks masses are $>1$ TeV.

\

Choosing $M_1$, $M_2$, $\mu$, $\tan\beta$ and the first generation slepton masses as in \eqref{GauLSP_parameters},\eqref{GauLSP_sfMass} permits to find a MSSM scenario with an approximately indistinguishable lower neutralino/chargino mass spectrum with that of NMSSM along all the $(\lambda,\,\kappa)$-plane.
The MSSM tree-level neutralino/chargino spectrum, is indeed given by:
\begin{center}
\begin{tabular}{|c|c|c|c|c|c|}\hline
$m_{\tilde{\chi}^0_1}$&$m_{\tilde{\chi}^0_2}$&$m_{\tilde{\chi}^0_3}$&$m_{\tilde{\chi}^0_4}$&$m_{\tilde{\chi}^{\pm}_1}$&$m_{\tilde{\chi}^{\pm}_2}$\\\hline
 99.5 GeV& 237.0 GeV   &510.1 GeV   &518.7 GeV  & 99.6 GeV & 518.7 GeV\\\hline
\end{tabular}             \end{center}
\normalsize

\begin{figure}[htb]\centering
\subfigure[]{\label{GAULSPN1mass}\includegraphics[width=.49\textwidth]{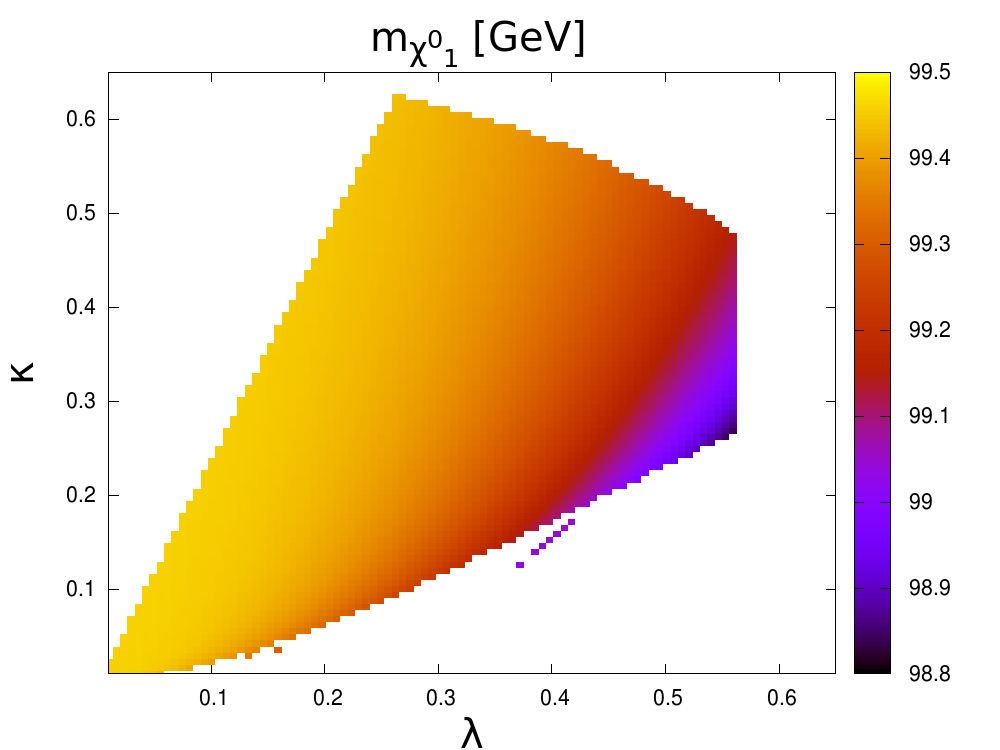}}
\subfigure[]{\label{GAULSPN15}\includegraphics[width=.49\textwidth]{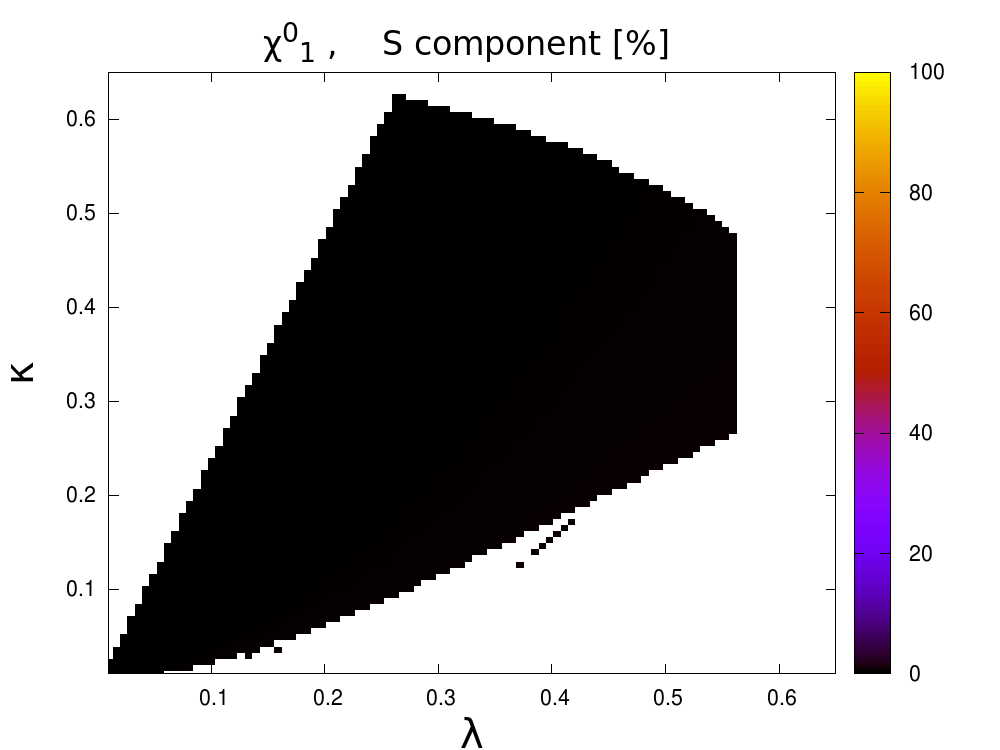}}
\caption{The light gaugino scenario: (a) the mass $m_{\tilde{\chi}^0_1}$, in GeV; (b) the $\tilde{S}$ component of $\tilde{\chi}^0_1$, in \%.\label{GAUN1plots}}
\end{figure}

In Figure \ref{GAUN1plots} one can see that the NMSSM $m_{\tilde{\chi}^0_1}$ is very close to 99.5 GeV from the MSSM and varies very mildly in the allowed in the $(\lambda,\,\kappa)$-plane since the singlino component in $\tilde{\chi}^0_1$ is approximately zero. Also, the production cross sections $\sigma(e^+e^-\rightarrow\tilde{\chi}^0_1\tilde{\chi}^0_2)$ are similar in the two models, while the $\sigma(e^+e^-\rightarrow\tilde{\chi}^+_1\tilde{\chi}^-_1)$  are exactly identical at the tree-level as explained in Subsection \ref{SubSecHIGLSP}.
\begin{center}
\begin{tabular}{|c||c|c|}\hline{\textbf{MSSM}}, $\sigma(e^+e^-\rightarrow\tilde{\chi}^0_1\tilde{\chi}^0_2)$  &$\sqrt{s}=350$~GeV & $\sqrt{s}=500$~GeV\\\hline\hline
$P=(-0.9,0.55)$&7.3 fb&113.4 fb\\\hline
$P=(0.9,-0.55)$&0.1 fb&1.8 fb\\\hline
\end{tabular} \end{center}\normalsize

\begin{center}
\begin{tabular}{|c||c|c|}\hline{\textbf{MSSM}}, $\sigma(e^+e^-\rightarrow\tilde{\chi}^+_1\tilde{\chi}^-_1)$  &$\sqrt{s}=350$~GeV & $\sqrt{s}=500$~GeV\\\hline\hline
$P=(-0.9,0.55)$&2692.1 fb&1252.6 fb\\\hline
$P=(0.9,-0.55)$&44.5 fb&19.4 fb\\\hline
\end{tabular} \end{center}

For the $\chi^2$-fit to the MSSM fit we use those cross sections that are large enough to be surely visible at the linear collider. We prefer to be conservative, so for $\tilde{\chi}^0_1\tilde{\chi}^0_2$ production we only use the $\sigma(e^+e^-\rightarrow\tilde{\chi}^0_1\tilde{\chi}^0_2)$ for $P=(0.9,-0.55)$ at 350 GeV, since the other cross sections are too small.

\begin{figure}[t!]
\begin{center}
 \includegraphics[width=3.0in]{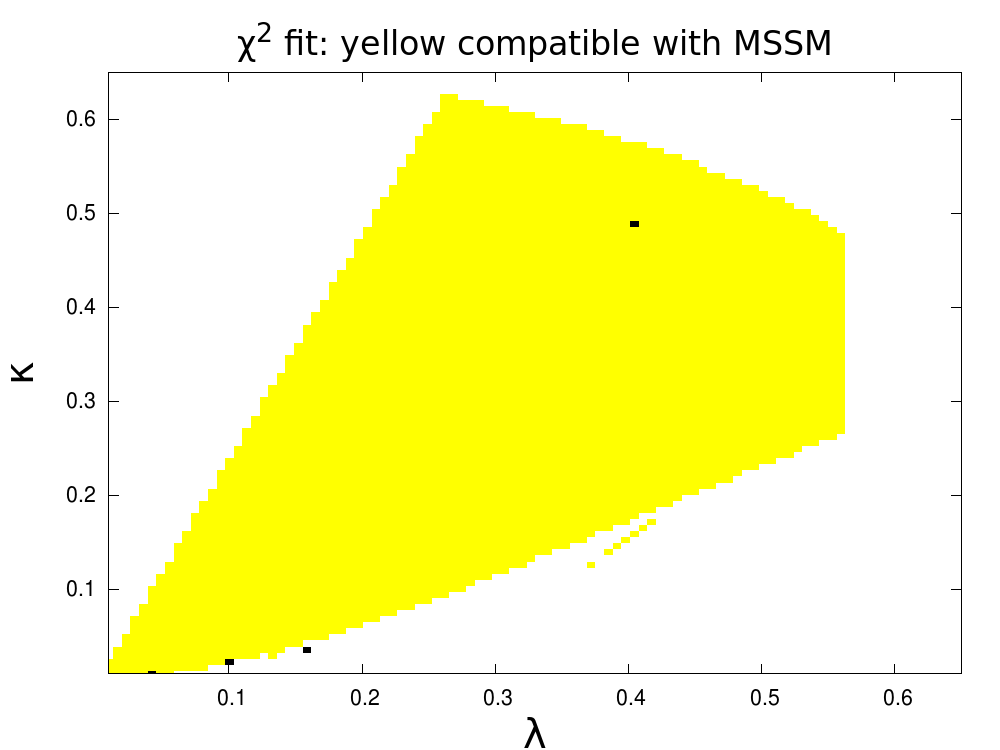}
 \caption{\label{GauLSP_Fit}Light gaugino scenario: fit to the MSSM. Yellow areas are compatible with the MSSM, black areas are not compatible.}
\end{center}
\end{figure}

Figure \ref{GauLSP_Fit} shows that our fit is not able to distinguish in this case between the two models.
The physically allowed region is indeed basically everywhere compatible with the MSSM.

\section{Conclusions and outlook}\label{Sec_Conclusions}
Supersymmetric models such as the MSSM and the NMSSM can lead to very similar light spectra. In case of SUSY discovery, methods to distinguish between the two models are needed. We addressed the study chargino and neutralino sectors alternatively to looking at the Higgs sector. It is proposed to distinguish between the MSSM and the NMSSM via $M_1$, $M_2$, $\mu=\mu_{\rm eff}$, $\tan\beta$ parameter reconstruction by fitting masses and cross sections from the chargino/neutralino sector. We have discussed two examples, the light higgsino and the light gaugino scenarios, where distinction could be possible.

\section*{Acknowledgements}

The authors are thankful to F.~Domingo, O.~St\aa{}l, L.~Zeune for useful discussions and support on \texttt{NMSSMTools} and \texttt{HiggsBounds/HiggsSignals}. The authors are also thankful to M.~Berggren for valuable insights on experimental issues.
S.~P. has been supported by DFG through the grant SFB 676 ``Particles, Strings, and the Early Universe''. This work has been partially supported by the MICINN, Spain, under contract FPA2010-17747; Consolider-Ingenio  CPAN CSD2007-00042. 
We thank as well the Comunidad de Madrid through Proyecto HEPHACOS S2009/ESP-1473 and the European Commission under contract PITN-GA-2009-237920.

\appendix

\bibliography{NMSSMvsMSSM_ILC}

\bibliographystyle{JHEP}

\end{document}